\documentclass[preprint2]{aastex}
\usepackage{graphicx}
\usepackage{natbib}
\usepackage{threeparttable}
\usepackage{url}
\usepackage{cite}

\slugcomment{Submitted to ApJ Letters on June 1, 2013}

\newcommand{\mos}{\,ms$^{-1}$}
\newcommand{\kms}{\,kms$^{-1}$}
\newcommand\msini{\ifmmode{{\mathrm M} \sin i}\else${{\mathrm M} \sin i}$\fi}

\shorttitle{Spin-orbit Misalignment of WASP-79}
\shortauthors{Addison et al.}

\begin{document}

\title{A Nearly Polar Orbit for the Extrasolar Hot Jupiter WASP-79b\footnote{Based on observations obtained at the Anglo-Australian Telescope, Siding Spring, Australia.}}

\author{B. C. Addison\altaffilmark{1,2}, C. G. Tinney\altaffilmark{1,2}, D. J. Wright\altaffilmark{1,2}, D. Bayliss\altaffilmark{3}, G. Zhou\altaffilmark{3}, J. D. Hartman\altaffilmark{4,5}, G. \'{A} Bakos\altaffilmark{4,5}, and B. Schmidt\altaffilmark{3}}

\email{b.addison@unsw.edu.au}

\altaffiltext{1}{Exoplanetary Science Group, School of Physics, University of New South Wales, 2052, Australia}
\altaffiltext{2}{Australian Centre of Astrobiology, University of New South Wales, 2052, Australia}
\altaffiltext{3}{Research School of Astronomy and Astrophysics, Australian National University, Canberra, ACT 2611, Australia}
\altaffiltext{4}{Department of Astrophysical Sciences, Princeton University, NJ 08544, USA}
\altaffiltext{5}{Harvard-Smithsonian Center for Astrophysics, Cambridge, MA 02138, USA}



\begin{abstract}
We report the measurement of a spin-orbit misalignment for WASP-79b, a recently discovered, bloated hot Jupiter from the WASP survey. Data were obtained using the CYCLOPS2 optical-fiber bundle and its simultaneous calibration system feeding the UCLES spectrograph on the Anglo-Australian Telescope. We have used the Rossiter-McLaughlin effect to determine the sky-projected spin-orbit angle to be $\lambda = -106^{+19}_{-13}$$^{\circ}$. This result indicates a significant misalignment between the spin axis of the host star and the orbital plane of the planet -- the planet being in a nearly polar orbit. WASP-79 is consistent with other stars that have $T_{eff}>6250$K and host hot Jupiters in spin-orbit misalignment.
\end{abstract}


\section{INTRODUCTION}
Over 850 exoplanets\footnote{\url{exoplanet.eu}, as of 2013 July. For ``discovery'' we adopt the requirement that a transit is both detected and confirmed by Doppler observations, and so do not include the several thousand planet candidates published by Kepler.} have been discovered to date using a variety of detection techniques. Doppler planet searches have historically been the most fruitful for finding planets (Bottom et al.~2013). In recent years, transit surveys have been leading the charge in discovering a host of new planets. These have been detected from ground based searches such as WASP (Pollacco et al.~2006), HAT-Net (Bakos et al.~2004), and HATSouth (Bakos et al.~2013), and more recently from space by surveys like Kepler (Borucki et al.~2010, Batalha et al.~2013). 

With so many planets now discovered, the focus has shifted to understanding their structure, composition, and other bulk properties so as to provide insights into the processes involved in planetary formation and migration (Santos 2008). One example of this is the combination of transit and radial velocity data to provide direct measurements of planetary bulk densities -- a physical measurement critical to the characterization of exoplanets composition and formation (Mordasini et al.~2012).

An additional probe of planetary formation and migration is provided by accurately measuring the sky-projected spin-orbit alignment (or obliquity) through spectroscopic measurements of the Rossiter-McLaughlin effect (first observed for eclipsing binary stars Rossiter~1924 and McLaughlin~1924; first observed for planets Queloz et al.~2000). This effect is caused by the modification of the stellar spectrum as a transiting planet occults a small region of the stellar disk of its host star, causing asymmetric distortions in the stellar line profiles that produce a radial velocity anomaly (Ohta et al.~2005). This effect is being detected for a growing number of planetary systems (see e.g. Albrecht et al.~2012b).

Planets are expected to form in the proto-stellar/proto-planetary disk that surrounds a protostar (e.g. Pollack et al.~1996). The dominant core-accretion paradigm for this formation process predicts that the stellar spin and planetary orbital axes should generally be aligned (e.g. Ohta et al.~2005; Winn et al.~2005) -- as in the case of our own solar system which is in alignment to within $6^{\circ}$ (Beck and Giles 2005). However, the large number of inward migrated Jovian exoplanets with orbital periods of only a few days seems to suggest that many exoplanetary systems have had a more complex formation history than the Solar System. Various planetary migration mechanisms -- including planet-planet scattering (e.g., Rasio \& Ford 1996; Weidenschilling \& Marzari 1996; Chatterjee et al. 2008), Kozai resonances (Kozai 1962; see, e.g., Naoz et al.~2011 for an application to highly inclined extrasolar planets), proto-planetary disks misalignments (Batygin~2012), or some combination of these processes (e.g. Nagasawa et al.~2008; Naoz et al.~2012; Hartman et al.~2012) -- have been proposed to explain misaligned systems though none have robustly predicted the misalignment of all observed systems.

In this letter, we present spectroscopic measurements obtained during the transit of WASP-79b, a recently discovered hot Jupiter from the WASP Southern Hemisphere transit survey (Smalley et al.~2012). We detect a clear radial velocity anomaly due to the Rossiter-McLaughlin effect and from these measurements determine that this system is significantly misaligned and lies in a nearly polar orbit.

\section{OBSERVATIONS \& ANALYSIS}
We obtained high-precision radial velocity observations of WASP-79 using the CYCLOPS2\footnote{\url{http://www.phys.unsw.edu.au/~cgt/CYCLOPS/CYCLOPS_2.html}} optical-fiber bundle feeding the UCLES echelle spectrograph on the 3.9m Anglo-Australian Telescope (AAT) at Siding Spring Observatory, Australia. CYCLOPS2 is a Cassegrain fiber-based integral field unit which reformats a $\sim 2.5^{\texttt{"}}$ diameter aperture into a pseudo-slit of dimensions equivalent to $0.6^{\texttt{"}}$ wide and $14.5^{\texttt{"}}$ long (Horton et al.~2012). It has replaced an earlier incarnation (CYCLOPS Classic)\footnote{\url{http://www.phys.unsw.edu.au/~cgt/CYCLOPS/CYCLOPS_Classic.html}} which had 15 fibers (of which 3 were inoperative) and $10\%$ lower throughput. CYCLOPS2 has 16 on-sky fibers, plus one fiber for simultaneous Thorium-Xenon lamp (ThXe) calibration. Each fiber delivers a spectral resolution of ${\lambda}/{\Delta}{\lambda} \approx 70,000$ over 19 echelle orders in the wavelength range of 4550--7350 \AA{}, when used with the UCLES spectrograph in its 79 line/mm grating configuration.

WASP-79 was observed continuously on the night of 2012 December 23, starting $\sim$20min before transit ingress and finishing $\sim$2h after egress. A total of 23 spectra were taken with 800s exposures and readout times of 120s for a cadence of 920s over a period of nearly 7h, with 14 exposures lying in the $\sim$3.8h transit duration. The overall observing conditions were very good with seeing ranging from $0.9^{\texttt{"}}$ to $1.2^{\texttt{"}}$ and mostly clear skies. The airmass at which WASP-79 was observed varied between 1.1 at the start of the night to $\sim1.0$ near mid transit and 1.7 at the end of the night. We obtained a S/N$=135$ per 2.5 pixel resolution element at $\lambda=5490$\AA{} (in total over all 16 fibers) when it was observed at an airmass of 1.1 and in $1^{\texttt{"}}$ seeing. To calibrate the observations, we used both a Thorium-Argon calibration lamp (ThAr) to illuminate all on-sky fibers and the ThXe lamp to illuminate the simultaneous calibration fiber. The simultaneous calibration system delivers two significant advantages. First, it eliminates the need for object observations to be bracketed by calibration images, providing an additional 240s on-sky per object exposure. Secondly, the error in the wavelength solution resulting from the interpolation of two bracketed calibration exposures is eliminated. 

The data were reduced using custom MATLAB routines developed by the authors. These use a master wavelength solution obtained from a ThAr image taken at the beginning of the night and simultaneous ThXe spectra taken during each object exposure. The simultaneous ThXe spectra are then used to make differential corrections to the master wavelength solution (Wright and Tinney, in prep.). Each of the 16 fibers, in each of the 18 useful orders, is used to estimate a radial velocity by cross-correlation with a spectrum of a bright template star (HD86264) of similar spectral type using the IRAF\footnote{IRAF is distributed by the National Optical Astronomy Observatories, which are operated by the Association of Universities for Research in Astronomy, Inc., under cooperative agreement with the National Science Foundation.} task, \textit{fxcor}. HD86264 was observed on the same night as WASP-79 at an airmass of 1.06 with a 1200s exposure delivering a S/N$=430$ per 2.5 pixel resolution element at $\lambda=5490$\AA{} (in total over all 16 fibers). \textit{Fxcor} implements the standard cross-correlation technique developed by Tonry and Davis (1979). Fitted Gaussians encompassing $\sim80\%$ of the peak in the cross-correlation function were used to compute a velocity (and associated uncertainty) for each of the $16\times18$ fiber-order (or ``fider") combinations. We experimented with a variety of templates for cross-correlation, including the highest signal-to-noise observation of WASP-79 and a synthetic spectrum. We found that the lowest inter-fider velocity scatter was obtained using the spectrum of HD86264. Weighted average velocities for each observation were determined by first clipping the fider velocities with Tonry \& Davis R numbers $<$ 10. The R number is the ratio height between the true peak and that of the average peak in the Gaussian fit to the cross-correlation function (Tonry and Davis 1979). We found that the fider velocities with an associated $R < 10$ were unreliable as the peak in cross-correlation function was not very Gaussian shaped (and thus could not be well fit) and not significantly above the noise in the rest of the cross-correlation function. After R clipping, a further $3\sigma$ clip was performed on the remaining fider velocities, from which a weighted mean was computed. Typically, a total of $<10\%$ of the velocities are rejected from clipping. Uncertainties for each of the weighted radial velocities were estimated from the weighted standard deviation of the fider velocity scatter. 

Table 1 shows our weighted radial velocities at each epoch, their uncertainties, and the total S/N over all 16 fibers per 2.5 pixel resolution element at a wavelength of $\lambda=5490$\AA{}. Also listed in Table 1 is a single velocity from Smalley et al. during the transit which has been phased and zero-point corrected to our data set.

\begin{table*}[ht]
\begin{threeparttable}[b]
\caption{Radial Velocities for WASP-79 (fiber and order averaged) with one previously published velocity during transit.}
\centering
\begin{tabular}{c c c c c c}
\hline\hline \\ [-2.0ex]
Time & RV & S/N at & Time & RV & S/N at \\
BJD-2400000 & (\mos) & $\lambda$=5490\AA{} & BJD-2400000 & (\mos) & $\lambda$=5490\AA{} \\ [0.5ex]
\hline \\ [-2.0ex]
55874.83089\tnote{a} & 5133 $\pm$ 49\tnote{a} & N/A\tnote{a} & 56285.07329 & 4929 $\pm$ 37 & 111 \\
56284.95307 & 4743 $\pm$ 42 & 132 & 56285.08398 & 4956 $\pm$ 36 & 114 \\
56284.96377 & 4751 $\pm$ 48 & 129 & 56285.09467 & 4829 $\pm$ 37 & 117 \\
56284.97446 & 4829 $\pm$ 38 & 133 & 56285.10537 & 4888 $\pm$ 31 & 128 \\
56284.98633 & 4881 $\pm$ 50 & 126 & 56285.11606 & 4713 $\pm$ 36 & 133 \\
56284.99701 & 4909 $\pm$ 41 & 129 & 56285.12675 & 4727 $\pm$ 36 & 127 \\
56285.00914 & 4969 $\pm$ 33 & 135 & 56285.13745 & 4677 $\pm$ 38 & 115 \\
56285.01984 & 4916 $\pm$ 35 & 133 & 56285.14814 & 4703 $\pm$ 38 & 107 \\
56285.03052 & 4914 $\pm$ 40 & 136 & 56285.15884 & 4638 $\pm$ 58 & 90 \\
56285.04122 & 4937 $\pm$ 39 & 107 & 56285.17011 & 4721 $\pm$ 39 & 102 \\
56285.05191 & 4912 $\pm$ 37 & 100 & 56285.18196 & 4705 $\pm$ 45 & 84 \\
56285.06260 & 5009 $\pm$ 37 & 109 & 56285.19381 & 4703 $\pm$ 39 & 91 \\
\hline \\ [-2.5ex]
\end{tabular}
\begin{tablenotes}
\item [a] \textit{Published in-transit radial velocity by Smalley et al. (2012).}
\end{tablenotes}
\end{threeparttable}%
\label{table:RVs}
\end{table*}
 
\begin{table*}[ht]
\begin{threeparttable}[b]
\caption{System parameters for WASP-79}
\centering
\begin{tabular}{l c c}
\hline\hline \\ [-2.0ex]
Parameter & Value (ms) & Value (non-ms) \\ [0.5ex]
\hline\hline \\ [-2.0ex]
{\textit{Parameters as given by Smalley et al. (2012)}} \\ [0.5ex]
{\textit{and used as priors in model}} \\
\hline \\ [-2.0ex]
Mid-transit epoch (2400000-HJD)\tnote{a}, $T_{0}$ & $56285.03589 \pm 0.00200$ & $56285.03739 \pm 0.00300$ \\
Orbital period\tnote{a}, $P$ & $3.6623817 \pm 0.0000050$ d & $3.6623866 \pm 0.0000085$ d \\
Semi-major axis, $a$ & $0.0539 \pm 0.0009$ AU & $0.0535 \pm 0.0008$ AU \\
Orbital inclination\tnote{a}, $i$ & $85.4 \pm 0.6^{\circ}$ & $83.3 \pm 0.5^{\circ}$ \\
Impact parameter\tnote{a}, $b$ & $0.570 \pm 0.052$ & $0.706 \pm 0.031$ \\
Transit depth\tnote{a}, $(R_{P}/R_{\star})^{2}$ & $0.01148 \pm 0.00051$ & $0.01268 \pm 0.00063$ \\
Orbital eccentricity\tnote{b}, $e$ & 0.0 (assumed) & 0.0 (assumed) \\
Argument of periastron, $\varpi$ & N/A ($e=0$) & N/A ($e=0$) \\
Stellar reflex velocity\tnote{b}, $K_{\star}$ & $0.0882 \pm 0.0078$ \kms\ & $0.0885 \pm 0.0077$ \kms\ \\
Stellar mass\tnote{b}, $M_{\star}$ & $1.56 \pm 0.09$ $M_{\odot}$ & $1.52 \pm 0.07$ $M_{\odot}$ \\
Stellar radius, $R_{\star}$ & $1.64 \pm 0.08$ $R_{\odot}$ & $1.91 \pm 0.09$ $R_{\odot}$ \\
Planet mass\tnote{b}, $M_{P}$ & $0.90 \pm 0.09$ $M_{J}$ & $0.90 \pm 0.08$ $M_{J}$ \\
Planet radius, $R_{P}$ & $1.70 \pm 0.11$ $R_{J}$ & $2.09 \pm 0.14$ $R_{J}$ \\
Stellar micro-turbulence\tnote{b}, $\xi_{t}$ & $1.3 \pm 0.1$\kms\ & $1.3 \pm 0.1$ \kms\ \\
Stellar macro-turbulence\tnote{b}, $v_{mac}$ & $6.4 \pm 0.3$ \kms\ & $6.4 \pm 0.3$ \kms\ \\
Stellar limb-darkening coefficient, $\mu$ & 0.606 (adopted) & 0.606 (adopted) \\
Velocity at published epoch $T_{P}$\tnote{b}, $V_{T_{P}}$ & $4.9875 \pm 0.0004$ \kms\  & $4.9875 \pm 0.0004$ \kms\ \\
RV offset\tnote{a}, $V_{d}$ & $0.2575$ \kms\ & $0.2575$ \kms\ \\ [0.5ex]
\hline \\ [-2.0ex]
\textit{Parameters determined from model fit using our velocities} \\ [0.5ex]
\hline \\ [-2.0ex]
Projected obliquity angle, $\lambda$ & $-106^{+19}_{-13}$ $^{\circ}$ & $-84^{+23}_{-30}$ $^{\circ}$ \\ [0.5ex]
Projected stellar rotation velocity, $v\sin i_{\star}$ & $17.5^{+3.1}_{-3.0}$ \kms\ & $16.0^{+3.7}_{-3.7}$ \kms\ \\ [0.5ex]
\hline \\ [-2.0ex]
\textit{Independent measurement of $v\sin i_{\star(Ind)}$ and} \\ [0.5ex]
{\textit{Smalley et al. (2012) $v\sin i_{\star(S)}$ published value}} \\ [0.5ex]
\hline \\ [-2.0ex]
Projected stellar rotation velocity, $v\sin i_{\star(Ind)}$ & $18.2 \pm {0.2}$ \kms\ & $18.2 \pm {0.2}$ \kms\ \\ [0.5ex]
Projected stellar rotation velocity, $v\sin i_{\star(S)}$ & $19.1 \pm {0.7}$ \kms\ & $19.1 \pm {0.7}$ \kms\ \\ [0.5ex]
\hline 
\end{tabular}
\begin{tablenotes}
\item [a] \textit{Parameters fixed to the indicated value for final fit, but allowed to vary (as described in \S{3}) for uncertainty estimation.}
\item [b] \textit{Parameters fixed at values given by Smalley et al. (2012).}
\end{tablenotes}
\end{threeparttable}%
\label{table:parameters}
\end{table*} 
 
\section{THE ROSSITER-McLAUGHLIN EFFECT} 
To determine the magnitude of the Rossiter-McLaughlin effect and accurately measure the spin-orbit alignment, we have developed a model we call the Exoplanetary Orbital Simulation and Analysis Model (ExOSAM). This model simulates the orbital position of a planet at the time of each observation (Prussing and Conway 2012). ExOSAM then computes the velocity from the motion of the star due to the orbiting planet and the in-transit lightcurve including the velocity anomaly due to the Rossiter-McLaughlin effect.

There are 13 parameters used by ExOSAM to compute the best fit model for the Rossiter-McLaughlin anomaly: two are free (the projected spin-orbit angle, $\lambda$, and projected stellar rotational velocity, $v\sin i_{\star}$); five are free but bounded to within the $2\sigma$ level uncertainties as given in Smalley et al.~(2012) (the planet-to-star radius ratio ($R_{p}/R_{\star}$), the orbital inclination angle ($i$), the orbital period ($P$), the mid transit time ($T_{0}$) at the epoch of observation, and the radial velocity offset ($V_{d}$) between the AAT and the Smalley et al. data sets (determined using observations from both data sets outside the transit event); and six are fixed and were adopted from Smalley et al. (planet-to-star mass ratio ($M_{p}/M_{\star}$), orbital eccentricity ($e$), argument of periastron ($\varpi$), the adopted linear limb-darkening coefficient ($\mu$), the micro-turbulence velocity ($\xi_{t}$), and the center-of-mass velocity ($V_{T_{P}}$) at published epoch $T_{P}$). We determined that the six fixed parameters negligibly contribute to the overall uncertainty in $\lambda$ and $v\sin i_{\star}$. These best fit parameters and their uncertainties are given in Table 2. 

The best-fitting values for $\lambda$ and $v\sin i_{\star}$ are derived using a grid search and minimizing $\chi^{2}$ between the observed radial velocities and modeled radial velocities. $\lambda$ and $v\sin i_{\star}$ were derived on a grid with intervals of $1.0^{\circ}$ in $\lambda$ and 0.1 \kms\ in $v\sin i_{\star}$ that was searched in the range $-75^{\circ}$ to $-130^{\circ}$ and 12.0\kms\ to 28.0\kms\ respectively. The $1\sigma$ confidence levels for these parameters were determined through the $\Delta \chi^{2}$ method (Bradt 2004) which is based on the probability distribution of $\chi^{2}$ as a function of the confidence level and degrees of freedom.

Various approaches for modeling the velocity anomaly ($\Delta v(t)$) caused by a planet as a function of its orbital parameters are available, including the first-moment approach (Ohta et al.~2005), stellar absorption line profile modeling approach (Hirano et al.~2011), and the forward-modeling approach (Winn et al.~2005). We have implemented the Hirano et al.~(2010) analytic solution given in Equation (1). It has been used in many studies (for example Bayliss et al.~2010) and was independently derived by Bou\'{e} et al.~(2013) specifically for the cross-correlation method of computing velocities.
\vspace{-0.5cm}

 \begin{equation}
 \resizebox{.53\vsize}{!}{$\Delta v(t)=-f(t)v_{p}(t)\left[\frac{2v^{2}_{0}+2({v\sin i_{\star}})^{2}}{2v^{2}_{0}+({v\sin i_{\star}})^{2}}\right]^{3/2}\left[1-\frac{v_{p}(t)^{2}}{2v^{2}_{0}+({v\sin i_{\star}})^{2}}\right]$}
 \end{equation}

The velocity anomaly as a function of time due to the Rossiter-McLaughlin effect is $ \Delta v(t)$, $f(t)$ is the flux and $v_{p}(t)$ is the ``sub-planet'' or line-of-sight velocity of the occulted region as a function of time, $v\sin i_{\star}$ is the projected stellar rotational velocity, and $v_{0}$ is the velocity width of the spectral lines from the occulted region due to mechanisms other than stellar rotation such as micro-turbulence and/or macro-turbulence.

The flux $f(t)$ is calculated by assuming a linear limb-darkening law for the stellar surface using the analytic equation given by Diaz-Cordoves et al.~(1995) and fixing an interpolated limb-darkening coefficient from look-up tables given in Diaz-Cordoves et al.~(1995) based on the assumed stellar parameters of WASP-79 from Smalley et al.~(2012). The sub-planet velocity $v_{p}(t)$ is calculated assuming that the rotation of the star is uniform (i.e. that it is not differentially rotating) with the angle $\lambda$ between the sky projected stellar spin axis and the planetary orbital angular momentum vector (Winn et al.~2005). The projected stellar rotational velocity $v\sin i_{\star}$ is a free parameter in our model.

\begin{figure}[h]
\centering
\includegraphics[width=1.0\linewidth]{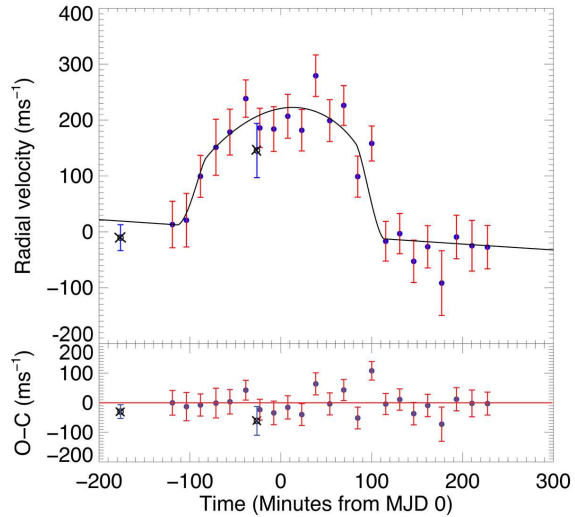}
\caption{Spectroscopic radial velocities of the WASP-79b transit. Velocity from just before, during, and after the transit are plotted as a function of time along with the best fitting model (for the main-sequence parameters) and corresponding residuals. The filled blue circles with red error bars are velocities we measured with our estimated uncertainty. The two black circles with an x and with blue error bars are previously published velocities by Smalley et al.~(2012) using their quoted uncertainties.}
\label{Figure_1_WASP-79b_transit}
\end{figure}

\begin{figure}[h]
\centering
\includegraphics[width=1.00\linewidth]{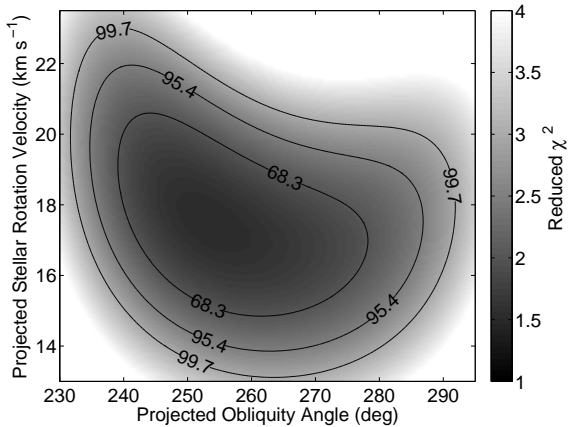}
\caption{The reduced $\chi^{2}$ as a function of $\lambda$ and $v\sin i_{\star}$ using the main sequence parameters of WASP-79. The contours show percentage confidence intervals.}
\label{Figure_2_contours}
\end{figure}

The best fit parameters and their uncertainties are given in Table 2. Smalley et al. were unable to uniquely determine the stellar parameters for WASP-79 from their photometric data, and derived two preferred solutions -- one with WASP-79 on the main sequence ($R_{\star}=1.64 \pm 0.08$ $R_{\odot}$) and one with it evolved just off the main sequence ($R_{\star}=1.91 \pm 0.09$ $R_{\odot}$). In addition, two different solutions for $R_{P}/R_{\star}$, $R_{\star}/a$, and $b$ are also given by Smalley et al. and are listed in Table 2. They suggest that the main sequence solution is the more likely one, nonetheless, we have performed our fit with both the main sequence and non-main sequence parameters for WASP-79. Figure \ref{Figure_1_WASP-79b_transit} shows the modeled velocity anomaly for the preferred main sequence solution with the observed velocities over-plotted. The filled blue circles with red error bars are velocities we measured, while the black circles with an x and with blue error bars are previously published velocities in Smalley et al.~(2012). 
The reduced $\chi^{2}$ as a function of $\lambda$ and $v\sin i_{\star}$ for the main sequence solution is shown in Figure \ref{Figure_2_contours}.

The Rossiter-McLaughlin effect is clearly detected as a positive ``hump-shaped'' anomaly (see Figure \ref{Figure_1_WASP-79b_transit}). The planet transits across only the blue-shifted hemisphere (i.e. the side rotating towards us) during the transit event and thus the star appears to be anomalously red-shifted. Based on the shape of the velocity anomaly, WASP-79b appears to be nearly in a polar orbit.

Our results for the projected spin-orbit alignment and stellar rotation velocity, using the main sequence parameters, are $\lambda = -106^{+19}_{-13}$\,$^{\circ}$ and $v\sin i_{\star} = 17.5^{+3.1}_{-3.0}$\kms. For the non-main sequence case, $\lambda = -84^{+23}_{-30}$\,$^{\circ}$ and $v\sin i_{\star} = 16.0^{+3.7}_{-3.7}$\kms. We obtain a much poorer fit for the non-main sequence solution, and a $\lambda$ estimate with a much larger uncertainty. Thus we agree with Smalley et al. that WASP-79 is more likely to be a main sequence star, however, high precision transit photometry is required in order to nail down this systems parameters. 

For the main sequence case of WASP-79, the $v\sin i_{\star}$ value we measured is consistent (within uncertainties) with Smalley et al.~(2012) value of $v\sin i_{\star}=19.1 \pm 0.7$\kms. As a further check of this value, we independently measured $v\sin i_{\star}$ using the highest signal-to-noise spectrum of WASP-79. This was done by fitting a rotationally broadened Gaussian (Gray 2005) to a least-squares deconvolution profile (Donati et al.~1997) obtained for each spectral order of WASP-79. The distribution of values along with an examination of the goodness of the fits was used to estimate the uncertainty for this $v\sin i_{\star}$ measurement (Gray 2005). This results in a value of $v\sin i_{\star} = 18.2 \pm 0.2$\kms\ -- again consistent with Smalley et al. and the Rossiter-McLaughlin measurements. 

\section{DISCUSSION}
As of July 2013, seventy-one\footnote{This study has made use of Ren\'{e} Heller's Holt-Rossiter-McLaughlin Encyclopaedia and was last updated on 2013 May. \url{www.aip.de/People/RHeller}} exoplanetary systems have measured spin-orbit alignments. Of these, 30 show substantial misalignments ($\lambda > \frac{\pi}{8} = 22.5^{\circ}$), 10 of which are in nearly polar orbits\footnote{We have adopted near-polar orbits as those with spin-orbit angles between $\frac{3\pi}{8} < \lambda < \frac{5\pi}{8}$ or $\frac{-3\pi}{8} > \lambda > \frac{-5\pi}{8}$ and retrograde orbits for spin-orbit angles between $\frac{5\pi}{8} \leq \lambda \leq \frac{11\pi}{8}$ or $\frac{-5\pi}{8} \geq \lambda \geq \frac{-11\pi}{8}$.} (including WASP-79b), and seven are in retrograde orbits\footnotemark[7]. With such a significant fraction of planets in spin-orbit misalignment, there is a clear need to understand the physical mechanisms that generates such high occurrence rates. Several mechanisms have been proposed based on apparent trends in planetary orbital obliquity. One such trend noted by Winn et al.~(2010a) is that hot stars with $T_{eff} > 6250$ K tend to host planets which are in spin-orbit misalignment. Winn et al.~(2010a) suggest that this can be understood if the convective layer in hot stars is too thin to effectively align the planet's orbital plane while cool stars have a thicker convective layer thus dampening orbital obliquities towards alignment in short time-scales. 

This conclusion has more recently been supported by the work of Albrecht et al. (2012b). After measuring the Rossiter-McLaughlin effect for 14 new systems and including the 39 previously published systems (at the time of their publication) in their analysis, they found a positive correlation between obliquity and stellar temperature (higher obliquity orbits around hotter stars) and a positive correlation for tidal dissipation timescales (higher obliquity orbits around stars with longer tidal dissipation timescales). They reason that hotter, more massive stars tend to have a thinner convective envelope as supported by stellar interior models (Pinsonneault et al.~2001), and that it is the convection envelope that is responsible for dampening the tidal energy and driving planets into alignment. This suggest that hotter stars are more likely to host planets in spin-orbit misalignment. They conclude by suggesting that the various mechanisms which caused Jupiter mass planets to migrate inwards in the first place also produced randomly distributed orbital obliquities (it is likely that hot Jupiters initially formed with low obliquity orbits as predicted by the core-accretion model as suggested by Ohta et al.~(2005) and Winn et al.~(2005)). After migration, systems with short tidal dampening timescales and strong tidal interactions quickly align their planets, while systems with longer tidal dampening timescales and/or weak tidal interactions will still display more random obliquity distributions long after migration.

Does WASP-79b align with the model of Albrecht et al.~(2012b)? WASP-79 has an effective temperature of $T_{eff}=6600\pm100$K, which is above the $T_{eff}>6250$K threshold claimed for planetary systems displaying more randomly distributed obliquities. Using either of the two methods presented by Albrecht et al., the tidal dissipation timescale for WASP-79 is found to be very long -- somewhere between $\tau_{mcz}=1.6\times10^{11}$yr to $\tau_{RA}=3.3\times10^{15}$yr (for an assumed main sequence age of WASP-79 between 0.5 and 3.5 Gyr). WASP-79, then, has a tidal dissipation timescale longer than 95\% of the systems examined by Albrecht et al., making it consistent with the observed trend of finding planets in high obliquity orbits in systems with long tidal-dampening timescales.

\section{CONCLUSION}
WASP-79b is in an orbit that is significantly misaligned with the projected rotational axis of its host star. Our most likely solution results in $\lambda = -106^{+19}_{-13}$ $^{\circ}$, though we cannot rule out a solution in which WASP-79 is an evolved star with $\lambda = -84^{+23}_{-30}$\,$^{\circ}$. This places the planet in a near-polar orbit. Conventional planetary formation models, such as core-accretion, do not predict Jovian type planets orbiting within 0.1AU from their host star or to be in highly misaligned orbits (Ohta et al.~2005; Winn et al.~2005). Yet the WASP-79 planetary system joins a growing list of known systems that are in significant spin-orbit misalignment (as shown in the study by Albrecht et al.~2012b). Additionally, planets that are thought to have undergone migration primarily due to the traditional Type 1 and Type 2 migration mechanisms (Lin et al.~1996) are predicted not to have their orbits significantly misaligned (Bate et al.~2010). 

A variety of mechanisms have been proposed to explain these systems (see for example Batygin~2012; Chatterjee et al.~2008; Naoz et al.~2011). However, it is only within the last year has there been a large enough sample of measured obliquities to begin looking for correlations and testing these mechanisms. Nonetheless, further expansion of the sample of planetary systems from which we can measure spin-orbit angles is needed to understand the mechanisms driving planetary migration and observed spin-orbit misalignments. Globally distributed ground-based transit searches such as HATSouth (Bakos et al.~2013) and new space based all-sky transit survey TESS (Deming et al.~2009) are set to deliver such samples before the decade is out, setting the scene for spin-orbit alignment measurements to play a key role in elucidating the complex formation and orbital evolution mechanisms of extra-solar planets.

The research work presented in the paper at UNSW has been supported by ARC Australian Professorial Fellowship grant DP0774000, ARC LIEF grant LE0989347 and ARC Super Science Fellowships FS100100046. Work at the Australian National University is supported by ARC Laureate Fellowship Grant FL0992131. We acknowledge the use of the SIMBAD database, operated at CDS, Strasbourg, France. 
 

\let\oldbibliography\thebibliography
\renewcommand{\thebibliography}[1]{%
  \oldbibliography{#1}%
  \setlength{\itemsep}{-3.5pt}%
}

\end{document}